\documentclass{article}
\usepackage{spconf,amsmath,epsfig,graphicx,amsfonts}

\title{RFI SUBSPACE SMEARING AND PROJECTION FOR ARRAY RADIO TELESCOPES}

\name{Gregory Hellbourg}
\address{Berkeley SETI Research Center - University of California - Berkeley, CA, USA}

\begin{document}

\maketitle

\begin{abstract}
Active Radio Frequency Interference (RFI) mitigation becomes a necessity for radio astronomy. The solution commonly applied by the community consists in monitoring the statistics of the received signal, and flag out the detected corrupted data. Subspace projection with array radio telescopes has been suggested as an alternative to data excision to avoid important losses of data and overcome its inherent ineffectiveness with continuous interference.

Spatial filtering relies on the estimation of the RFI spatial contribution, and the projection of the subspace spanned by the RFI out of the observed data vector space. To perform well, the dimensionality of the RFI subspace is constrained.

RFI subspace estimation techniques assume the source of RFI to be spatially stationary over the sample covariance matrix evaluation. When the relative movement between the telescope and the interferer becomes significant, the RFI subspace gets smeared over the whole data vector space. The subspace projection can then no longer be applied without affecting the source of interest recovery.

This paper addresses the effect of RFI subspace smearing on the subspace projection approach, and suggests an alternative technique based on a covariance matrix subtraction, improving the performance of spatial filtering in the case of high subspace smearing.
\end{abstract}


\section{Introduction}

Radio Frequency Interference (RFI) is a threat to radio astronomy, despite the establishment of protective measures such as Radio Quiet Zones, protected frequency bandwidths and on-site equipment shielding \cite{davis2010spectrum}.

As astronomers are mostly interested in recovering time and frequency signatures of cosmic sources, the exploitation of RFI spatial features to perform their active mitigation appears as a natural alternative to standard RFI cancellation methods applied in the time and frequency domains \cite{ghose1996interference}. Moreover, array radio telescopes are suitable for the implementation of spatial filters \cite{1502901}.

Synthesis imaging exploits the correlation between the elements of an interferometer to sample the astronomical image in the Fourier domain \cite{perley1989synthesis}. An orthogonal projection is applied to the array covariance matrix after estimating the subspace spanned by an interferer within the observed data vector space \cite{246805}. This approach is referred to as subspace projection.

Several techniques for estimating the RFI subspace have been proposed \cite{6334141}. A major limitation to these techniques comes from the accuracy of the estimation of the covariance matrix itself, based on multiple realizations of the array output vector. This paper investigates the impact of the relative movement between a telescope and the source of RFI over the covariance matrix estimation duration on the RFI subspace estimate. It is shown that the RFI subspace may asymptotically span the whole data vector space, limiting the performance of subspace projection at recovereing the source of interest. In this case, a covariance matrix subtraction approach is introduced as a substitute.

Section \ref{sec:datamodel} presents the data model assumed throughout the paper, and defines the orthogonal projection operator. Section \ref{sec:smearing} considers the Sample Covariance Matrix as an estimator of the data covariance matrix and introduces the concept of RFI subspace smearing. Section \ref{sec:subtraction} presents the covariance matrices subtraction approach, improving the performance of spatial filtering in case of high subspace smearing. Section \ref{sec:conclusion} concludes the paper.

\section{Data model and subspace projection}
\label{sec:datamodel}

The output of an array radio telescope made of $M$ elements contains the contributions of astronomical sources, RFI and the system noise. Astronomical sources are commonly neglected due to their weakness\footnote{Astronomical sources are recovered after long data integration (of the order of hours).}. We further assume that only one RFI is present within the considered frequency channel\footnote{Typical processing frequency bandwidths span $\approx$ 1 kHz.}. The gain-calibrated and narrow band output vector is then expressed as:

\begin{equation}
\mathbf{x}(t) = \mathbf{r}(t) + \mathbf{n}(t)
\label{eq:time_model}
\end{equation}

where:

\begin{itemize}
\item $\mathbf{x}(t)$ is the $M \times 1$ array output vector,
\item $\mathbf{r}(t) = r(t) \mathbf{a}_r(t)$ is the RFI contribution with:
\begin{itemize}
\item $r(t)$ the RFI signal, and
\item $\mathbf{a}_r(t)$ the $M \times 1$ RFI spatial signature vector (SSV) made of phase shifts induced by the RFI propagation among the array elements
\end{itemize}
\item $\mathbf{n}(t)$ is the $M \times 1$ system noise vector
\end{itemize}

The interferer is considered to be a continuous waveform $r(t) = \sigma_r e^{i\left(\omega t + \phi\right)}$, $\left(\sigma_r,\omega,\phi\right) \in \mathcal{R}^3$. Without loss of generality, we assume $\left\|\mathbf{a}_r(t)\right\| = 1$ with $\left\|.\right\|$ the Euclidean vector norm. The system noise is assumed to be circular, temporally and spatially white, stationary over a short period, independently and identically distributed (i.i.d.) with $\mathbf{n}(t) \sim \mathcal{NC}\left(\mathbf{0}, \sigma_n^2 \mathbf{I}\right)$, $\mathbf{I}=\mathbf{I}_M$ being the $M \times M$ identity matrix and $\mathcal{NC}\left(\mathbf{\mu}, \mathbf{R}\right)$ the multivariate complex normal distribution with mean vector $\mathbf{\mu}$ and covariance matrix $\mathbf{R}$.

The array covariance matrix is defined as:

\begin{align}
\mathbf{R}(t,\tau) &= \mathbb{E}\left\{ \mathbf{x}(t)\mathbf{x}(t-\tau)^H \right\}\nonumber\\
&= \underbrace{\sigma_r^2 e^{i \omega \tau}\mathbf{a}_r(t)\mathbf{a}_r(t-\tau)^H}_{\mathbf{R}_{\text{RFI}}(t,\tau)} + \underbrace{\delta(\tau)\sigma_n^2\mathbf{I}}_{\mathbf{R}_{\text{noise}}(t,\tau)}
\label{eq:covar}
\end{align}
with $\mathbb{E}\left\{.\right\}$ the expectation operator, $\left(.\right)^H$ the conjugate transpose operator and $\delta(\tau)$ the Kronecker delta such that:

\begin{equation*}
\delta(\tau) = 
\begin{cases} 
1 & \text{if} \  \tau = 0\\
0 & \text{else}
\end{cases}
\end{equation*}

The subspace projection approach consists in estimating the RFI subspace (generated by $\mathbf{a}_r$) and projecting it out of the observed data vector space evaluated at $\tau=0$ with the following orthogonal projection operator \cite{boonstra2005radio}:

\begin{equation}
\mathbf{P}(t) = \mathbf{I} - \mathbf{a}_r(t) \left(\mathbf{a}_r^H(t) \mathbf{a}_r(t)\right)^{-1} \mathbf{a}_r^H(t)
\end{equation}

This operator is either applied to the array output vector ($\mathbf{x}_P(t) = \mathbf{P}\mathbf{x}(t)$) or to the covariance matrix ($\mathbf{R}_P(t,\tau=0) = \mathbf{P}\mathbf{R}(t,\tau=0)\mathbf{P}^H$). Following the data model \ref{eq:covar}, the dominant eigen vector of $\mathbf{R}(t,\tau)$ is an estimate $\widehat{\mathbf{a}}_r$ of the RFI SSV $\mathbf{a}_r$, but other estimation approaches based on statistical properties of the RFI have been considered \cite{6334141}.

The RFI subspace is not necessarily limited to a 1-dimensional subspace. Multiple sources of RFI can co-exist at the same frequency (e.g. with a navigation satellites constellation) and increase the RFI subspace dimensionality. In this case, the subspace dimension has to be estimated \cite{1164557,1502941} and a set of independent vectors spanning the same subspace are then chosen to be projected out of the data vector space.

\section{Subspace smearing and limits of subspace projection}
\label{sec:smearing}

\subsection{Sample covariance matrix}

The array covariance matrix is unknown and needs to be estimated with the data provided by the telescope. A classic estimator of $\mathbf{R}(t,\tau)$, broadly implemented on telescope arrays systems, is the Sample Covariance Matrix (SCM) evaluated over $N$ data samples and defined as:

\begin{equation}
\widehat{\mathbf{R}}(\tau) = \frac{1}{N}\sum_{n=0}^{N-1}{\mathbf{x}(n.T_s)\mathbf{x}(n.T_s-\tau)^H}
\end{equation}
where $T_s$ is the signal sampling period.

The SCM is the maximum likelihood estimator of the covariance matrix in the Gaussian model. In the spatially stationary case (i.e. $\mathbf{a}_r(n.T_s) = \mathbf{a}_r$, $\forall n$) and based on the model \ref{eq:time_model}, the statistical properties of this estimator are given by (details omitted due to space limitation):

\begin{align}
\mathbb{E}\left\{\widehat{\mathbf{R}}(\tau)\right\} &= \frac{1}{N}\sum_{n=0}^{N-1}{\mathbb{E}\left\{\mathbf{x}(n.T_s)\mathbf{x}(n.T_s-\tau)^H\right\}}\nonumber\\
&= \mathbf{R}(\tau)
\end{align}
and
\begin{align}
&\text{var}\left\{\text{vec}\left(\widehat{\mathbf{R}}\right)\right\} = \mathbb{E}\left\{ \text{vec}\left(\widehat{\mathbf{R}}\right) \text{vec}\left(\widehat{\mathbf{R}}\right)^H \right\}\nonumber\\
&= \frac{1}{N \rho} [ \sigma_r^4 \left( \mathbf{I} \otimes \mathbf{a}_r\mathbf{a}_r^H + \bar{\mathbf{a}}_r\bar{\mathbf{a}}_r^H \otimes \mathbf{I}\right) + \sigma_r^2\sigma_n^2\mathbf{I}_{2M} -\nonumber\\
& \delta(\tau)\sigma_r^2\sigma_n^2 \left( \text{vec}\left(\mathbf{I}\right)\text{vec}\left(\mathbf{I}\right)^H  + \mathbf{I}_{2M}\right) ]
\label{eq:varSCM}
\end{align}
with $\rho = \frac{\sigma_r^2}{\sigma_n^2}$ the Signal-to-Noise Ratio (SNR), $\text{vec}\left(.\right)$ the vectorization operator and $\otimes$ the Kronecker product operator (time indexes suppressed for simplicity).

\begin{figure}
\centering
\includegraphics[width=0.4\textwidth]{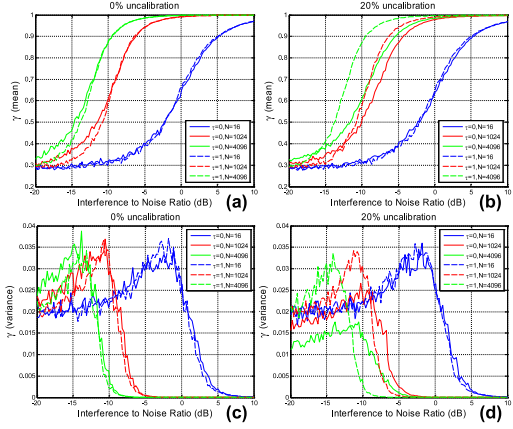}
\caption{Performance of the RFI SSV estimation based on the dominant eigenvalue of $\widehat{\mathbf{R}}(\tau)$ for $\tau=0$ and $\tau=1$, function of the Interference-to-Noise Ratio and the number of samples N over which the SCM is evaluated. Performance given in terms of the mean and variance of the dot product $\gamma$ between the estimated and the true RFI SSV over 512 Monte-Carlo trials. (a) and (b) : mean of $\gamma$ with 0\% and 20\% gain-uncalibration, respectively. (c) and (d) : variance of $\gamma$ with 0\% and 20\% gain-uncalibration, respectively.}
\label{fig:analysis}
\end{figure}

Equation \ref{eq:varSCM} shows the dependence of the variance of the SCM on the SNR and the number of samples $N$. As a consequence, the estimate of the RFI SSV based on the dominant eigen value gets also affected by this dependence. Figure \ref{fig:analysis} shows the performance of this estimator in terms of the mean value and variance of the dot product $\gamma = \frac{\left|\mathbf{a}_r^H \mathbf{u}_1\right|}{\left\|\mathbf{a}_r\right\|\left\|\mathbf{u}_1\right\|}$ between $\mathbf{a}_r$, the true SSV, and $\mathbf{u}_1$, the dominant eigen vector of $\widehat{\mathbf{R}}(\tau)$ for $\tau=0$ and $\tau=1$ sample, with $\left|.\right|$ standing for the absolute value. $\gamma$ follows $0 \leq \gamma \leq 1$, $\gamma = 1$ implying a collinearity between the vectors involved, and $\gamma = 0$ their orthogonality. The gain-uncalibrated scenario, where $\mathbf{x}_u(n) = \text{diag}\left(\mathbf{u}\right)\cdot\mathbf{x}(n)$ (with $\text{diag}\left(.\right)$ the vector to diagonal matrix transformation operator and $\mathbf{u}$ is a vector with uniformly distributed entries $\mathbf{u}_k \sim \mathcal{U}(1-\delta,1+\delta)$, $k=1..M$), is compared to the calibrated calibrated scenario. A 20\% gain fluctuation over the array elements is considered (i.e. $\delta = .1$). The results have been averaged over 512 Monte-Carlo trials.

This analysis leads to the following comments:

\begin{itemize}
\item A large number of samples is required for performing an accurate RFI subspace estimation, and hence an accurate RFI subspace projection.
\item The gain-uncalibration of the array deteriorates the quality of the RFI subspace estimation for low SNR when $\tau=0$. Large gain fluctuations over the array elements uncorrelates the RFI SSV estimate from the dominant eigenvalue.
\item Estimating the RFI SSV with the SCM evaluated at $\tau \neq 0$ however improves the estimation accuracy in the gain-uncalibrated scenario as the noise contribution gets canceled for large $N$.
\end{itemize}

\subsection{RFI subspace smearing}

As the number of samples increases, the RFI SSV accuracy increases but the spatial stationarity of this vector can no longer be neglected. Considering the source of RFI entering the array through the same side lobe over the SCM short evaluation time, the $1^{\text{st}}$ order approximation of the phases variations for each component of its SSV is linear and given by (for the $k^{\text{th}}$ entry):

\begin{equation}
\mathbf{a}_{r_k}(n.T_s) = \frac{1}{\sqrt{M}}e^{i(\alpha_k n.T_s + \phi_k)}
\label{eq:movingRFI}
\end{equation}
with $\left(\alpha_k,\phi_k\right) \in \mathcal{R}^2$. We have, for short $\tau$, $\mathbf{a}_r(n.T_s)\approx\mathbf{a}_r(n.T_s-\tau)$. The $(k,l)^{\text{th}}$ element of the RFI SCM is then given by:

\begin{align}
\widehat{\mathbf{R}}_{\text{RFI}_{k,l}}(\tau) &= \frac{1}{N} \sum_{n=0}^{N-1}r(n.T_s)\bar{r}(n.T_s-\tau)\cdot\nonumber\\
&\mathbf{a}_{r_k}(n.T_s)\mathbf{a}_{r_l}^H(n.T_s)\nonumber\\
&= \sigma_r^2\frac{e^{i \omega \tau}}{MN} \sum_{n=0}^{N-1} e^{i\left(\alpha_k-\alpha_l\right)n.T_s}e^{i\left(\phi_k-\phi_l\right)}\nonumber\\
&= \sigma_r^2\frac{e^{i \omega \tau}e^{i\left(\phi_k-\phi_l\right)}}{MN}\frac{1 - e^{i\left(\alpha_k-\alpha_l\right)N}}{1 - e^{i\left(\alpha_k-\alpha_l\right)}}
\label{eq:coeff_rfi_scm}
\end{align}
with $\bar{(.)}$ the complex conjugate operator. From \ref{eq:coeff_rfi_scm} we have:

\begin{equation}
\lim_{N \to \infty}\widehat{\mathbf{R}}_{\text{RFI}_{k,l}}(\tau) \approx
\begin{cases} 
0 & \text{if} \ \alpha_k \neq \alpha_l \\
\sigma_r^2\frac{e^{i \omega \tau}e^{i\left(\phi_k-\phi_l\right)}}{M} & \text{else}
\end{cases}
\label{eq:smearing}
\end{equation}

This result shows that the dimension of the RFI subspace increases with the number of samples. Particularly, if $\forall (k,l) \in [1..M]^2$, $k \neq l$, we have $\alpha_k \neq \alpha_l$, then:

\begin{equation}
\lim_{N \to \infty} \widehat{\mathbf{R}}_{\text{RFI}}(\tau) \approx \sigma_r^2\frac{e^{i \omega \tau}}{M} \mathbf{I}
\label{eq:full_smearing}
\end{equation}
and the RFI spans the whole observed data vector space. The situation where $\alpha_k = \alpha_l$, $\forall (k,l) \in [1..M]^2$, while the relative movement between the telescope and the RFI source is non-negligible only occurs if the $M$ array elements are co-located.

\section{Subtraction approach}
\label{sec:subtraction}

The increase of the number of samples over which the SCM is evaluated, as well as the amplitude of the relative motion between the interferer and the telescope, leads to an increase of the dimension of the RFI subspace as shown in equation \ref{eq:smearing}. When a significant ratio of the RFI power is concentrated over only a few dimensions of the data vector space (e.g. for short SCM estimation time or small ``RFI - telescope'' relative movement amplitude), a low rank approximation of the RFI subspace can be performed \cite{scharf1991statistical} and the orthogonal projector can still be applied. Depending on the data model parameters, the RFI rejection can then be evaluated according to the number of dimensions of the RFI subspace projected out.

When this low rank approximation is no longer valid (equation \ref{eq:full_smearing} for instance), the RFI subspace projection is not relevant anymore. In this case, an alternative solution consists in subtracting the RFI covariance matrix from the array covariance matrix. We consider now the following extended data model taking in account the astronomical sources contributions:

\begin{equation}
\mathbf{x}(t) = \mathbf{r}(t) + \sum_{k=1}^{N_c} c_k(t) \mathbf{a}_{c_k}(t)  + \mathbf{n}(t)
\label{eq:time_model_astro}
\end{equation}
where $\mathbf{a}_{c_k}$ is the SSV of the $k^{th}$ astronomical source with waveform $c_k(t)$. Due to the underlying physical processes leading to the generation of $c_k(t)$, these signals are assumed temporally white, stationary, i.i.d with $c_k(t) \sim \mathcal{NC}(0,\sigma_{c_k}^2)$. In general, $\sigma_{c_k}^2 < \sigma_n^2$, $\forall k \in [1..N_c]$.

The subtraction approach consists in subtracting the RFI contribution from the array covariance matrix:

\begin{align}
\mathbf{R}_{\text{sub}}(t,\tau) &= \mathbf{R}(t,\tau) - \mathbf{R}_{\text{RFI}}(t,\tau)\nonumber\\
&= \delta(\tau) \sum_{k=1}^{N_c} \sigma_{c_k}^2 \mathbf{a}_{c_k}(t) \mathbf{a}_{c_k}^H(t-\tau) + \mathbf{R}_{\text{noise}}(t,\tau)
\end{align}

Due to the whiteness of both astronomical sources and system noise, the covariance matrix evaluated at a time lag $\tau \neq 0$ approximates the spatial structure of $\mathbf{R}_{\text{RFI}}(t,\tau)$ for short $\tau$. Without a priori knowledge regarding the modulation parameters of the RFI, the corrected array SCM follows, for short $\tau$:

\begin{equation}
\widehat{\mathbf{R}}_{\text{sub}}(\tau) = \widehat{\mathbf{R}}(\tau=0) - \xi_0 \widehat{\mathbf{R}}(\tau \neq 0)
\end{equation}

with $\xi_0$ expressed as:

\begin{equation}
\xi_0 = \underset{\xi}{\text{argmin}} \left\| \widehat{\mathbf{R}}(\tau=0) - \xi \widehat{\mathbf{R}}(\tau \neq 0) \right\|^2_{\text{F}}
\label{eq:optimization}
\end{equation}
$\left\|.\right\|_F$ standing for the Frobenius norm operator.

Nulling the first derivative of equation \ref{eq:optimization} leads to:

\begin{equation}
\xi_0 = \frac{\text{tr}\left( \widehat{\mathbf{R}}(\tau \neq 0)^H \widehat{\mathbf{R}}(\tau = 0) \right)}{\text{tr}\left(\widehat{\mathbf{R}}(\tau \neq 0)^H \widehat{\mathbf{R}}(\tau \neq 0)\right)}
\end{equation}
with $\text{tr}\left(.\right)$ the trace operator.

\begin{figure}
\centering
\includegraphics[width=0.4\textwidth]{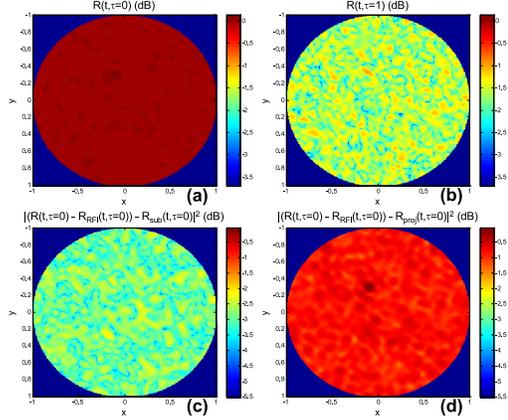}
\caption{Comparison between subtraction approach and orthogonal projection in case of RFI subspace smearing. M = 100 antennas. (a) Raw data with astronomical source (SNR = -5 dB) and moving RFI (INR = +10 dB). (b) Time lagged data, asymptotically no contribution from cosmic source and system noise. (c) Residual error between RFI free data and subtraction approach. (d) Residual error between RFI free data and orthogonal projection approach.}
\label{fig:sub}
\end{figure}

To illustrate the improvement brought by the subtraction approach, figure \ref{fig:sub} shows an example of simulated astronomical data involving one cosmic source (SNR = -5 dB) and a moving interferer (INR = +10 dB) following the structure presented in equation \ref{eq:movingRFI} with $\alpha_k \sim \mathcal{N}\left(0,0.1\right)$. The SCM are evaluated over $N =$ 1024 samples, and the array is made of $M =$ 100 antennas randomly distributed (maximum baseline length $\approx$ 15 $\lambda$). Figure \ref{fig:sub}.(a) shows the raw data. The cosmic source, located at coordinates [-0.3,-0.1], is barely visible due to the weak SNR. The RFI is moving over the integration time, and can therefore not be seen on the map. Figure \ref{fig:sub}.(b) corresponds to the data evaluated with a $\tau = 1$ time lag. This map is supposed to contain much less contribution of the source and the noise, the RFI remains invisible. Figures \ref{fig:sub}.(c) and \ref{fig:sub}.(d) correspond to the mean squared error between the RFI free data and the corrected data after the subtraction approach and the orthogonal projection approach, respectively. As expected, the error is lower with the subtraction approach due to the data vector space deformation induced by the orthogonal projector \cite{991140}.

\section{Conclusion}
\label{sec:conclusion}

This paper presents the limit of the subspace projection approach for interference mitigation when the relative movement between the telescope and a source of RFI is not negligible. The evaluation of the Sample Covariance Matrix (SCM) over a large time window improves the covariance estimation variance, but increases the RFI subspace smearing. The dimensionality of this subspace increases according to the number of samples over which the SCM is evaluated and eventually spans the whole data vector space.

The use of a time-lagged SCM is suggested for isolating the RFI contribution. This matrix is then subtracted from the classic SCM, allowing a better recovery of astronomical sources and system noise. This approach does not require any a priori knowledge regarding the data model parameters, but can be improved with exploiting, when available, information regarding the interference modulation scheme or cyclostationariry for instance \cite{hellbourg2012cyclostationary}.

The amount of time lags over which the RFI SCM is evaluated is further constrained as the auto-correlation of a modulated signal has a limited support in the time lags domain. Another limitation comes from the structure of some cosmic sources being temporally non white. Also, the RFI covariance matrix estimation remains dependent of the number of samples over which it is evaluated, and the interference-to-noise ratio.

\bibliographystyle{IEEEbib}
\bibliography{refs}

\end{document}